\documentclass[conference]{IEEEtran}
\IEEEoverridecommandlockouts
\usepackage{amsmath,amssymb,amsfonts}
\usepackage{algorithmic}
\usepackage{graphicx}
\usepackage{textcomp}
\usepackage{xcolor}
\usepackage{soul} 
\usepackage{array}
\usepackage{multirow}
\usepackage[hidelinks]{hyperref}
\usepackage{dblfloatfix}    
\usepackage[sort&compress,numbers]{natbib} 
\usepackage{float}
\usepackage{csquotes}
\usepackage{gensymb}
\usepackage{amsmath,texlogos}

\usepackage{flexisym} 

\newcommand{\R}{\mathbb{R}} 

\DeclareUnicodeCharacter{2212}{\textendash}

 \def\BibTeX{{\rm B\kern-.05em{\sc i\kern-.025em b}\kern-.08em
    T\kern-.1667em\lower.7ex\hbox{E}\kern-.125emX}}

\begin{document}

\title{Time-Synchronized State Estimation Using Graph Neural Networks in Presence of Topology Changes}

\author{\IEEEauthorblockN{Shiva Moshtagh, \textit{Student Member, IEEE}, Anwarul Islam Sifat, \textit{Member, IEEE},\\Behrouz Azimian, \textit{Student Member, IEEE}, and Anamitra Pal, \textit{Senior Member, IEEE} \\
\textit{School of Electrical, Computer, and Energy Engineering} \\
\textit{Arizona State University},
Tempe, AZ, USA\\
\href{mailto:smoshta1@asu.edu}{smoshta1@asu.edu}, \href{mailto:anwarul.sifat@asu.edu}{anwarul.sifat@asu.edu}, \href{mailto:bazimian@asu.edu}{bazimian@asu.edu}, and \href{mailto:anamitra.pal@asu.edu}{anamitra.pal@asu.edu} 
}


\thanks{This work was supported in part by the Department of Energy (DOE) under the grants DE-EE0009355 and DE-AR-0001001, and the National Science Foundation (NSF) under the grant ECCS-2145063.}

}

\maketitle

\begin{abstract}
Recently, there has been a major emphasis on developing data-driven approaches involving machine learning (ML) for high-speed static state estimation (SE) in power systems.
The emphasis stems from the ability of ML to overcome difficulties associated with model-based approaches, such as
handling of non-Gaussian measurement noise.
However, topology changes pose a stiff challenge for performing ML-based SE because the training and test environments become different when such changes occur.
This paper circumvents this challenge by formulating a graph neural network (GNN)-based time-synchronized state estimator that considers the physical connections of the power system during the training itself.
The results obtained using the IEEE 118-bus system indicate that the GNN-based state estimator outperforms both the model-based linear state estimator and a data-driven deep neural network-based state estimator in the presence of non-Gaussian measurement noise and topology changes, respectively.

\end{abstract}

\vspace{0.5em}
\begin{IEEEkeywords}
Graph neural network (GNN), Machine learning (ML), State estimation (SE), and Topology change.
\end{IEEEkeywords}

\section{Introduction}
\label{Intro}
Static state estimation (SE) is performed in modern energy management systems to enhance situational awareness of power system operators \cite{monticelli2000electric}.
In static SE, the state refers to the voltage magnitudes and angles of all the vertices of the power system graph, where the vertices are the buses located inside substations.
In the past decade, intelligent electronic devices (IEDs) called phasor measurement units (PMUs) have been placed in bulk inside the substations to the effect that many power systems are completely, and often redundantly, observed by these devices (e.g., see Fig. 1 of \cite{zhang2017design}).
PMUs rely on the global positioning system to provide time-stamped 
measurements.
Consequently, the estimation performed using these measurements is synchronized in time; a phenomenon that is commonly referred to as \textit{time-synchronized estimation} in the power system literature.

Linear state estimation (LSE) is a type of time-synchronized estimation in power systems in which the static states are estimated solely based on PMU data.
Outputs of linear state estimators are preferred over those of supervisory control and data acquisition (SCADA)-based state estimators because of the former's high speed (typically 30 samples/second) and time-synchronized nature.
When a system is fully observed by PMUs, LSE is performed by solving the following equation:

\begin{equation}\label{eq:LSE-model}
z = Hx + e
\end{equation}

In \eqref{eq:LSE-model}, $z \in {\mathbb{R}}^{m}$ is the measurement vector containing $m$ phasor measurements, $x \in {\mathbb{R}}^{s}$ is the state variable vector to be estimated, where $s$ is the number of unknown states, and $e \in {\mathbb{R}}^m$ is the measurement noise vector. Matrix $H \in {\mathbb{R}}^{m \times s}$ in \eqref{eq:LSE-model} is the Jacobian measurement matrix that relates the measurements with the states. It is defined in \cite{phadke2017phasor} as:
\begin{equation}\label{eq:H}
    H = 
\begin{bmatrix}
  II\\
  YA_1+Y_s \\
\end{bmatrix}
\end{equation}
where, $II$ and $A_1$ are the voltage measurement-bus incidence matrix and the current measurement-bus incidence matrix, respectively, while $Y_s$ and $Y$ are the series and shunt admittance matrices. The admittance matrices are a function of the network topology, and are calculated from line parameters, which are assumed to be known \textit{apriori}. 
The LSE solution is typically obtained by minimizing the modeling error in the least squares sense \cite{abur2004power} as shown below:

\begin{equation}\label{eq:LSE-sol}
\hat{x}_{\mathrm{LSE}}=(H^{\top}H)^{-1}H^{\top}z
\end{equation}

In \eqref{eq:LSE-sol}, $\hat{x}_{\mathrm{LSE}}$ is the solution to the maximum likelihood estimation problem under Gaussian noise environments \cite{varghese2022transmission}.
However, it has recently been demonstrated that PMU noise exhibits non-Gaussian characteristics 
\cite{ahmad2019statistical,salls2021statistical}.
The primary basis for the non-Gaussian noise in PMU measurements is the instrumentation system of this IED, which comprises instrument transformers, attenuators, burdens, and cables.
As the performance of linear estimation (LSE is an example of linear estimation) can deteriorate in the presence of non-Gaussian measurement noise \cite{principe2010information},
it is important to investigate strategies that can perform time-synchronized SE when non-Gaussian noise is present in the PMU measurements.


Recently, we have demonstrated the ability of machine learning (ML) to perform high-speed time-synchronized SE in distribution systems in presence of non-Gaussian noise in synchrophasor measurements \cite{azimian2021time}. In \cite{azimian2021time}, the ML model that was developed was a \textit{regular} (also called vanilla) deep neural network (DNN).
However, a regular DNN is not explicitly aware of the topology of the power system. This implies that it may struggle to maintain its accuracy when the topology of the system changes \cite{azimian2022state}.
In \cite{gotti2021deep}, a TI DNN model was created to identify the topology of transmission systems at high speeds. Such a model could be used as a precursor to a regular DNN created for performing SE in presence of topology changes.
However, 
placing two DNNs in a series may deteriorate estimation performance as the errors of the first one will impinge on the second.

In this paper, we investigate if an alternate DNN architecture, namely, graph neural networks (GNNs), can continue
to perform accurate and consistent SE after a topology change \textit{and without any subsequent re-training}.
GNN is an ML model that
works on non-Euclidean
data defined as a graph \cite{hamilton2020graph}. 
The inputs of a GNN are the node feature and adjacency matrices which contain the node and connectivity information of the graph, respectively.
By employing graph-structured data, one can incorporate physical information of the system into the learning process of the GNN \cite{bronstein2017geometric}.
Hence, the GNN's learning process includes knowledge of the data (features) as well as the topology of the 
system. This makes GNNs distinct from other ML models and more suited for handling topology changes.

Because of the unique characteristics of GNNs, 
they are gaining popularity in solving different types of power system problems.
In the context of transmission system SE, \cite{yang2022data} combined a Gauss-Newton solver with a GNN model to perform SE using only SCADA data. 
Ref. \cite{wu2022state} performed GNN-based SE by combining SCADA and PMU data to create a fully-observed node feature matrix, which is a requirement for GNNs \cite{chen2020learning}.
However, a purely SCADA-based state estimator or a SCADA-PMU hybrid state estimator does not have the advantages of a PMU-only state estimator, namely, \textit{time-synchronized outputs at high speeds}.
Recently, a PMU-only framework has been presented in \cite{kundacina2022state}, which used factor graph-based GNN models for performing SE.
However, the measurement noise was assumed to be Gaussian in \cite{kundacina2022state}, and the impact of topology changes was not analyzed.
In this paper, similar to \cite{kundacina2022state}, we first place PMUs optimally to obtain a fully-observed node feature matrix; however, we do not create a separate
factor-graph (the reason is given in Section \ref{GNN}). 
Then, we build a GNN in which the physical connections of the power system are embedded using multiple convolutional layers and important node features are extracted via an attention layer, to perform high-speed time-synchronized SE.
Finally, we demonstrate the robustness of the proposed GNN-based state estimator compared with the state-of-the-art for high-speed SE, namely, model-based LSE and regular DNN-based SE, in terms of handling non-Gaussian measurement noise and topology changes, respectively.

The rest of the paper is structured as follows. Section \ref{GNN} presents the architecture of the proposed GNN. The results obtained when this GNN was used for performing SE in the IEEE 118-bus system are presented in Section \ref{Results}. The concluding remarks are provided in Section \ref{Conclusions}.


\section{Graph Neural Network-based State Estimation}\label{GNN}


An electric grid comprising $n$ buses can be modeled as a graph, $\mathcal{G}=(\mathcal{V},\mathcal{E})$, where $\mathcal{V}=\{v_1,\hdots, v_n\}$ is the set of nodes (vertices), and $\mathcal{E}=\{(v_i,v_j)\}\subseteq\mathcal{V}\times\mathcal{V}$ is the set of edges (transmission lines and transformers). Each node represents a feature vector that is included in the node feature matrix, $X \in \R^{n \times d}$, where $d$ denotes the number of features for each node.
Node connectivity is defined in the form of the adjacency matrix,  $\mathrm{A} \in \R^{n \times n}$, whose elements indicate whether pairs of nodes are connected in the graph:

\begin{equation}\label{eq:adjacency}
\mathrm{A_{ij}} = \begin{cases}
  1 \hspace{10pt} (v_i,v_j) \in \mathcal{E}\\
  0 \hspace{10pt} \text{ otherwise}
\end{cases}
\end{equation}

If PMUs are only placed on some of the nodes of the system, there will be missing features in 
$X$.
We overcome this problem by 
placing PMUs such that all the nodes either have PMUs on them or are adjacent to at least one node that has a PMU on it.
Such optimal locations can be determined from prior literature on optimal PMU placement for LSE (e.g., \cite{pal2013pmu}).
Note that loss of PMU data can result in the feature matrix becoming unobservable. 
Contrary to creating a separate factor-graph to overcome this issue as done in \cite{kundacina2022state}, we suggest detecting and replacing bad/missing PMU data \textit{before} it enters the GNN as already done for regular DNNs in \cite{mestav2019bayesian}.

\begin{figure}[b]
\centering
\includegraphics[width=0.489\textwidth]{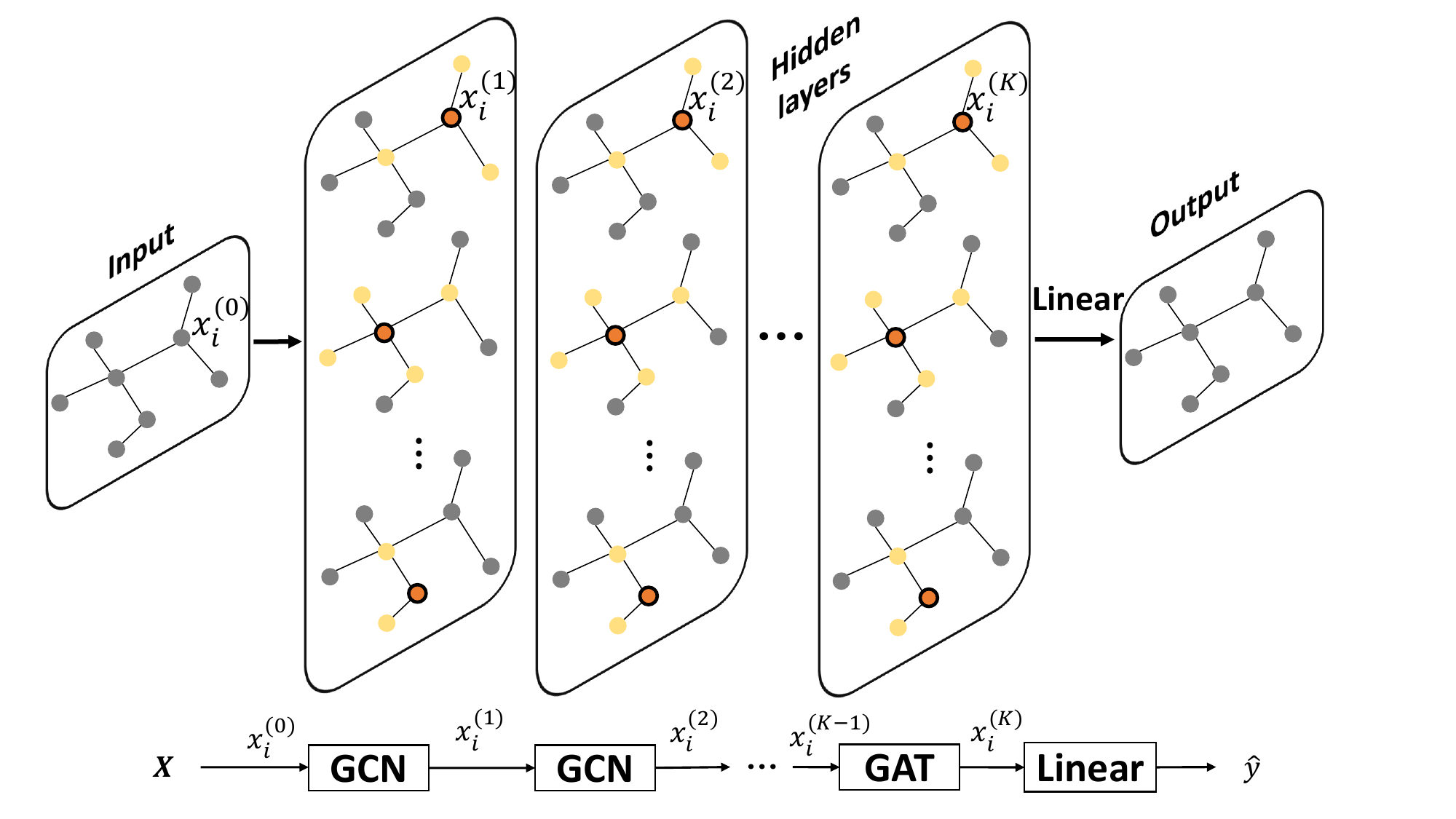}
\vspace{-1.5em}
\caption{Proposed GNN model for SE - \textbf{Input}: graph-based data with initial feature values obtained from PMU measurements. \textbf{Hidden layers}: features of every node (orange nodes) are updated in parallel by aggregating neighboring nodes' information (yellow nodes) in each hidden layer. \textbf{Output}: the final graph representation with state estimates that are obtained by applying a linear transformation to the final hidden layer.}
\label{figure}
\end{figure}

To embed the physical connections of the power system into the GNN model, we use a recursive neighborhood aggregation scheme or message-passing procedure, which occurs inside the hidden layers, as illustrated in Fig. \ref{figure}.
Message-passing aggregates neighboring
nodes’ information based on an aggregation function for every node in the graph.
The type of aggregation used for message-passing determines how messages of neighboring nodes are 
combined.
The two aggregation types used in the proposed GNN model are described below. 

\textit{Graph Convolutional Network (GCN):}
GCN is used to exploit topology information in the system graph by aggregating weighted features in the neighborhood of every node.
It is mathematically defined as:
\begin{equation}\label{eq:GCN1}
x^{(k)}_i=\sigma \left( \sum_{j\in\mathcal{N}(i)\cup\{i\} }W^{(k)}\frac{x^{(k-1)}_j}{\sqrt{|\mathcal{N}(i)||\mathcal{N}(j)|}} \right ) 
\end{equation}
where, $x^{(k)}_i \in \mathbb{R}^{d'}$ and $x^{(k-1)}_j \in \mathbb{R}^d$ are feature vectors of nodes $i$ and $j$ at layers $k$ and $k-1$, respectively, $\sigma$ is the rectified linear-unit (ReLU) activation function, $W \in \mathbb{R}^{d' \times d}$ is the learnable weight matrix, $|\mathcal{N}(i)|$ and $|\mathcal{N}(j)|$ are the number of nodes in the neighborhood of nodes $i$ and $j$, respectively, and $k=1,2,\hdots, \mathrm{K}$ refers to the different hidden layers of the GNN.
The matrix, $W$, transforms input node features to higher-level features 
to capture complex relationships and patterns in the graph data. 
In each iteration of \eqref{eq:GCN1}, features of every node are updated by propagating the weighted features of its neighboring nodes 
to capture the physical connections of the power system graph. 

Lastly, \eqref{eq:GCN1} can be expressed in matrix form as:
\begin{equation}\label{eq:GCN2}
X^{(k)}=\sigma(\Tilde{\mathrm{A}}X^{(k-1)}{W^\top}^{(k)})
\end{equation}
where $\Tilde{\mathrm{A}}=D^{-1/2}\hat{\mathrm{A}}D^{-1/2}$ is the normalized version of the adjacency matrix with inserted self loops (denoted by $\hat{\mathrm{A}}$), and $D_{ii}=\sum_{j} \hat{\mathrm{A}}_{ij}$ is the diagonal degree matrix of $\hat{\mathrm{A}}$. The self loops are inserted into the conventional adjacency matrix, $\mathrm{A}$, by adding an identity matrix, $\mathrm{I}$, of the same dimension with it; i.e., $\hat{\mathrm{A}}=\mathrm{A}+\mathrm{I}$.


\textit{Graph Attention Network (GAT):}
GAT is used to add importance to specific messages from neighbors of each node in the graph, as opposed to GCN, where all the messages are treated as equally important. 
A GAT is formulated as:

\begin{equation}\label{eq:GAT1}
x^{(k)}_i=\sigma \left(\sum_{j\in\mathcal{N}(i)\cup\{i\}}\alpha_{ij}W^{(k)}x^{(k-1)}_j \right)
\end{equation}
where, $\alpha_{ij}$ is the level of importance of neighboring node $i$'s message to node $j$, and is calculated as:

\begin{equation}\label{eq:}
\alpha_{ij}=\frac{\mathrm{exp}(e_{ij})}{\sum_{j\in\mathcal{N}(i)\cup\{i\}}\mathrm{exp}(e_{ij})}
\end{equation}
where the attention coefficient, $e_{ij}$, is given by,

\begin{equation}\label{eq:7}
e_{ij}=\sigma_{\mathrm{leaky}}(\mathrm{a}^\top[W^{(k)}x^{(k-1)}_i \mathbin\Vert W^{(k)}x^{(k-1)}_j])
\end{equation}

In \eqref{eq:7}, $\mathbin\Vert$ refers to the concatenation operation, and $\sigma_{\text{leaky}}$ is the leaky-ReLU activation function. GAT adds the learnable parameter $\mathrm{a} \in \R^{2d'}$
to learn the attention (importance) across pairs of nodes during the training process. 
The key benefit of using GAT is its ability to assign different levels of importance to different neighbors and obtain a better feature representation for each node, thereby improving estimation performance.

Eq. \eqref{eq:GAT1} can now be written in matrix form as:
\begin{equation}\label{eq:GAT2}
X^{(k)}=\sigma(\mathrm{C}X^{(k-1)}{W^\top}^{(k)})
\end{equation}
where $\mathrm{C} \in R^{n \times n}$ is the coefficient matrix whose elements are $\alpha_{ij}$ if $(v_i,v_j) \in \mathcal{E}$, and 0 otherwise. 

In the last step, the state variables are predicted by applying a linear transformation to the node feature matrix of the $\mathrm{K}^{th}$ layer.
This is mathematically described by: 
\begin{equation}\label{eq:SE}
\hat{y_{i}}=X^{(\mathrm{K})}W^{'}+b^{'}
\end{equation}
where, $X^{(\mathrm{K})}$ is the node feature matrix of the $\mathrm{K}^{th}$ layer, $W^{'}$ and $b^{'}$ are the learnable parameters of the linear layer,  and $\hat{y_{i}}$ is the state estimate obtained using the GNN for the $i^{th}$ train/test sample.


The next section demonstrates how the
framework mentioned above enables
the proposed GNN-based state estimator to perform robust time-synchronized SE under different system and sensing conditions.

\section{Simulation Results}\label{Results}

We now showcase the performance of the proposed GNN-based SE approach compared to LSE and a regular DNN-based SE for the IEEE 118-bus system. 
Specifically, we verify the robustness of the proposed approach to non-Gaussian noise in the PMU measurements and topology changes in the system.

\subsection{Simulation Setup}\label{Simulation Setup}
To train the proposed GNN-based state estimator, we created 28,000 graph samples to represent different operating conditions (OCs). Each graph sample denotes a realistic load consumption scenario 
and its corresponding state variables.
The node feature matrix, $X$, 
comprises noisy voltage magnitude and phase angles of every bus, and is created from measurements coming from PMUs placed at optimal locations. 
32 such locations were identified, which is also sufficient for performing LSE in this system \cite{pal2013pmu}.
Synchrophasor measurements were synthetically generated by solving an AC power flow using MATPOWER, followed by a corruption of the data via addition of noise. The noise model used in this study follows a 1\% total vector error (TVE) non-Gaussian distribution according to IEEE standard \cite{6111219}.
Note that $X$ is specific to each graph sample, while the adjacency matrix, $\mathrm{A}$, is fixed for all the training samples.

The matrices $X$ and $\mathrm{A}$ are the inputs to the GNN model, and the state variables (noise-free voltage magnitudes and angles) are the outputs.
The GNN structure comprises five GCN and one GAT layer (i.e., a total of six hidden layers), where the GCN layers extract the topology-based information (physical connections), while the GAT layer learns the importance of neighborhood information within the graph. 
The number of epochs, learning rate, and optimizer are 1,000, 0.001, and Adam, respectively.
The test data comprises 4,000 distinct graph samples 
for every topology (see Section \ref{Numerical Results3} regarding how the different topologies were created).

We used PyTorch and TensorFlow libraries in Python to implement the proposed GNN and the benchmark DNN models, respectively. The LSE was implemented in MATLAB. All simulations were carried out on a high-performance computer with 256 GB RAM, Intel Xeon 6246R CPU at 3.40 GHz, and Nvidia Quadro RTX 5000 16 GB GPU.


\subsection{Robustness to Non-Gaussian Measurement Noise}\label{Numerical Results}

\subsubsection{Comparison with LSE}\label{Numerical Results1}
We first compare the performance of the proposed GNN-based state estimator with a linear state estimator in presence of Gaussian and non-Gaussian noise in PMU measurements. 
The non-Gaussian noise was modeled as a two-component Gaussian mixture model, whose mean, standard deviation, and weights for magnitudes and angles are $(-0.4\%, 0.6\%)$ and $(-0.2\degree,0.3\degree)$, $(0.25\%, 0.25\%)$ and $(0.12\degree, 0.12\degree)$ and $(0.4, 0.6)$, respectively \cite{varghese2022transmission}.
The results of the comparison, shown in Table \ref{table1}, indicate 
that GNN-based SE performs better than LSE in terms of the mean absolute percentage error (MAPE) for magnitudes and the mean absolute error (MAE) for angles for both noise models. 
The mathematical expressions of MAPE and MAE are,

\begin{subequations}
\label{subeqn:list}
\begin{align}
\mathrm{MAPE} = \frac{1}{\mathrm{N}}\sum_{i=1}^{\mathrm{N}} \left | \frac{y_i-\hat{y_i}}{y_i} \right|
 \label{subeqn:first} \\
    \mathrm{MAE} = \frac{1}{\mathrm{N}}\sum_{i=1}^{\mathrm{N}}|y_i-\hat{y_i}|
 \label{subeqn:second}
\end{align}
\end{subequations}
where, $\mathrm{N}$ is the number of graph samples, and $y_i$ and $\hat{y_i}$ denote the actual and estimated states, respectively.
The deterioration in the state estimates, when the noise model changes from Gaussian to non-Gaussian, is particularly acute for the LSE as the error almost doubles (see Table \ref{table1}).
However, there is only a minuscule difference in the performance of the proposed GNN-based SE when the noise model changes. 

\begin{table}[hb]
\renewcommand\thetable{I}
\centering
\caption{Estimation Error Comparison between LSE and GNN-based SE for IEEE 118-bus system}\label{table1}
\resizebox{\columnwidth}{!}{
\begin{tabular}{ccc|cc}
\cline{2-5}
\multirow{3}{*}{\textbf{}} & \multicolumn{2}{c|}{ \rule{0pt}{1\normalbaselineskip}\textbf{Gaussian Noise}}& \multicolumn{2}{c}{\textbf{Non-Gaussian Noise}} \\[2pt] \cline{2-5}
& \rule{0pt}{1\normalbaselineskip} \textbf{MAE} & \textbf{MAPE}  & \textbf{MAE}  & \textbf{MAPE} \\ 
& \textbf{\begin{tabular}[c]{@{}c@{}}Phase angle\\(degrees)\end{tabular}} & \textbf{\begin{tabular}[c]{@{}c@{}}Magnitude\\ (\%)\end{tabular}} & \textbf{\begin{tabular}[c]{@{}c@{}}Phase angle\\ (degrees)\end{tabular}} & \textbf{\begin{tabular}[c]{@{}c@{}}Magnitude\\  (\%)\end{tabular}} \\[2pt]  \hline
\rule{0pt}{1\normalbaselineskip}\textbf{LSE}   &  0.149 & 0.270 &  0.246 &  0.492 \\[2pt]  \hline
\rule{0pt}{1\normalbaselineskip} \textbf{GNN-SE} &  0.080 &  0.047 &   0.095   &  0.054   \\[2pt]  \hline
\end{tabular}
}
\end{table}

\subsubsection{Comparison with Regular DNN-based SE}\label{Numerical Results2}

Next, we compare the performance of the proposed GNN-based state estimator with a regular DNN-based state estimator in presence of non-Gaussian noise in the PMU measurements.
The noise model is identical to the one used in Section \ref{Numerical Results1}.
The DNN has six hidden layers each having 200 neurons, with the other hyperparameters (number of epochs, learning rate, and optimizer) being same as that of the GNN.
The results of this comparison are shown in Table \ref{table2}.
In this table, along with MAE and MAPE, we also compute the $\mathrm{R^2}$-score (called R-squared) for both the state estimators.
This criterion measures the contribution percentage of the input data to track variation in the output labels, and is mathematically defined as:

\begin{equation}\label{eq:R2} \tag{10}
\mathrm{R^2} = 1- \frac{\sum_{i=1}^{\mathrm{N}}(y_i-\hat{y_i})^2}{\sum_{i=1}^{\mathrm{N}}(y_i-\bar{y_i})^2}
\end{equation}
where, $\bar{y_i}$
denotes the mean value across all the output labels.
From the higher $\mathrm{R^2}$-scores of the proposed GNN-based state estimator, we can infer that it can track more of the variations occurring in the output labels in comparison to the DNN-based state estimator (see Table \ref{table2}).
This improved performance 
is due to the fact that the GAT layer is able to provide contextual feature representation for each node by focusing on the most important parts of the input data and fading out the rest utilizing the ML concept of attention mechanism.

\begin{table}[hb]
\renewcommand\thetable{II}
\caption{Estimation Error Comparison between Regular DNN-based SE and GNN-based SE for IEEE 118-bus system}\label{table2}
\resizebox{\columnwidth}{!}{%
\begin{tabular}{ccccc}
\cline{2-5}
 & \rule{0pt}{1\normalbaselineskip}\textbf{MAE} & \textbf{MAPE} & \multicolumn{2}{c}{\textbf{R-Squared}}\\[2pt] 
 & \multicolumn{1}{c}{\begin{tabular}[c]{@{}c@{}}\textbf{Phase angle}\\ {\textbf{(degrees)}}\end{tabular}} & \begin{tabular}[c]{@{}c@{}}\textbf{Magnitude}\\ {\textbf{(\%)}}\end{tabular} & 
 \multicolumn{1}{c}{\begin{tabular}[c]{@{}c@{}}\textbf{Phase angle}\\ {\textbf{}}\end{tabular}} & \begin{tabular}[c]{@{}c@{}}\textbf{Magnitude}\\ {\textbf{}}\end{tabular} \\[2pt]
\hline
\rule{0pt}{1\normalbaselineskip}\textbf{DNN-SE} & \multicolumn{1}{c}{0.125} & 0.096 & \multicolumn{1}{c}{0.94} &0.65\\[2pt]
\hline
\rule{0pt}{1\normalbaselineskip}\textbf{GNN-SE} & \multicolumn{1}{c}{0.095} &  0.054 & \multicolumn{1}{c}{0.97} &0.82\\[2pt]
\hline
\end{tabular}%
}
\end{table}

\subsection{Robustness to Topology Changes}\label{Numerical Results3}
Topology change is a common occurrence in the power system, and depending on the impact that the change has on the OC of the system, it can cause minor or major deterioration in the performance of ML-based estimation.
Hence, in this subsection, we investigate the sensitivity of the proposed GNN-based SE framework to topology changes in the IEEE 118-bus system.
The linear state estimator and DNN-based SE used in the previous sections provide the baselines.
Both ML models (DNN-SE and GNN-SE) were trained for the nominal (base) topology, but tested on data related to off-nominal topologies when a single-line outage occurs by feeding this data to a pre-trained model. For the LSE, which does not consist of any training, the model of the system ($H$) presented in \eqref{eq:LSE-model} and \eqref{eq:H} was kept fixed, while the measurements, $z$, in \eqref{eq:LSE-model} came from different topologies associated with single-line outages, as explained below.

\begin{table*}[ht]
\renewcommand\thetable{III}
    \caption{Comparison of MAE and MAPE between LSE, DNN-SE, and proposed GNN-SE models for five different topologies of the IEEE 118-bus system associated with outages of the lines with the lowest power flow (T\textprime) and outages of the lines with the highest power flow (T) considering non-Gaussian measurement
noise}\label{table3}
\resizebox{\textwidth}{!}{%
\begin{tabular}{ccccc || ccccc}
\hline
\multirow{2}{*}{\textbf{Topology}} & \multirow{2}{*}{\textbf{\begin{tabular}[c]{@{}c@{}}From-To buses of\\ the removed line\end{tabular}}} & \multirow{2}{*}{\textbf{Model}} & \multirow{2}{*}{\textbf{\begin{tabular}[c]{@{}c@{}}MAE Phase angle\\ (degrees)\end{tabular}}} & \multirow{2}{*}{\textbf{\begin{tabular}[c]{@{}c@{}}MAPE Magnitude\\ (\%)\end{tabular}}} & \multirow{2}{*}{\textbf{Topology}} & \multirow{2}{*}{\textbf{\begin{tabular}[c]{@{}c@{}}From-To buses of\\ the removed line\end{tabular}}} & \multirow{2}{*}{\textbf{Model}} & \multirow{2}{*}{\textbf{\begin{tabular}[c]{@{}c@{}}MAE Phase angle\\ (degrees)\end{tabular}}} & \multirow{2}{*}{\textbf{\begin{tabular}[c]{@{}c@{}}MAPE Magnitude\\ (\%)\end{tabular}}} \\
 &  &  &  &  &  &  &  &  &  \\ \hline
\multirow{3}{*}{\textbf{T\textprime1}} & \multirow{3}{*}{24 - 70} & LSE & 0.246 & 0.670 & \multirow{3}{*}{\textbf{T1}} & \multirow{3}{*}{8 - 5} & LSE & 0.433 & 0.679 \\
 &  & DNN-SE & 0.138 & 0.101 &  &  & DNN-SE & 3.435 & 0.749 \\
 &  & \textbf{GNN-SE} & 0.136 & 0.081 &  &  & \textbf{GNN-SE} & 0.591 & 0.220 \\ \hline
\multirow{3}{*}{\textbf{T\textprime2}} & \multirow{3}{*}{56 - 58} & LSE & 0.249 & 0.682 & \multirow{3}{*}{\textbf{T2}} & \multirow{3}{*}{30 - 17} & LSE & 0.329 & 0.681 \\
 &  & DNN-SE & 0.141 & 0.108 &  &  & DNN-SE & 3.224 & 0.707 \\
 &  & \textbf{GNN-SE} & 0.151 & 0.099 &  &  & \textbf{GNN-SE} & 0.277 & 0.177 \\ \hline
\multirow{3}{*}{\textbf{T\textprime3}} & \multirow{3}{*}{100 - 101} & LSE & 0.269 & 0.741 & \multirow{3}{*}{\textbf{T3}} & \multirow{3}{*}{26 - 30} & LSE & 0.246 & 0.671 \\
 &  & DNN-SE & 0.132 & 0.119 &  &  & DNN-SE & 1.382 & 0.644 \\
 &  & \textbf{GNN-SE} & 0.135 & 0.093 &  &  & \textbf{GNN-SE} & 0.396 & 0.215 \\ \hline
\multirow{3}{*}{\textbf{T\textprime4}} & \multirow{3}{*}{14 - 15} & LSE & 0.246 & 0.670 & \multirow{3}{*}{\textbf{T4}} & \multirow{3}{*}{38 - 37} & LSE & 0.335 & 0.702 \\
 &  & DNN-SE & 0.131 & 0.099 &  &  & DNN-SE & 1.804 & 0.713 \\
 &  & \textbf{GNN-SE} & 0.137 & 0.086 &  &  & \textbf{GNN-SE} & 0.301 & 0.322 \\ \hline
\multirow{3}{*}{\textbf{T\textprime5}} & \multirow{3}{*}{32 - 113} & LSE & 0.246 & 0.670 & \multirow{3}{*}{\textbf{T5}} & \multirow{3}{*}{64 - 65} & LSE & 0.325 & 0.673 \\
 &  & DNN-SE & 0.143 & 0.119 &  &  & DNN-SE & 1.338 & 0.386 \\
 &  & \textbf{GNN-SE} & 0.142 & 0.098 &  &  & \textbf{GNN-SE} & 0.284 & 0.146 \\ \hline
\end{tabular}%
}
\end{table*}

We can create new (off-nominal) topologies by removing one line at a time from the base topology. 
To create diverse off-nominal topologies, we removed those lines that had the lowest or the highest powers flowing through them in the base topology, while ensuring that the system remained connected. 
The results of this attempt for three models LSE, DNN-based SE, and GNN-based SE are summarized in Table \ref{table3} for the outages of five lines of the IEEE 118-bus system that have the lowest (${\textbf{T\textprime}}$) and highest (${\textbf{T}}$) power flowing through them.

Table \ref{table3} compares the robustness of LSE, DNN-SE, and the proposed GNN-SE to topology changes in terms of MAE and MAPE considering non-Gaussian measurement noise.
It can be observed from Table \ref{table3} that for the off-nominal topologies, ${\textbf{T\textprime}}$, the results were similar for the two ML models (DNN-SE and GNN-SE) as well as LSE. This is due to the fact that the alteration in the base OC, after those lines that carried the lowest powers was removed, was negligible. However, for the off-nominal topologies denoted by ${\textbf{T}}$, the GNN-based state estimator consistently outperformed the DNN-based state estimator as well as LSE.
This highlights the robustness of the proposed GNN-SE approach to topology changes because of its topology-aware structure.

\begin{figure*}[ht]
\centering
\includegraphics[width=1\textwidth]{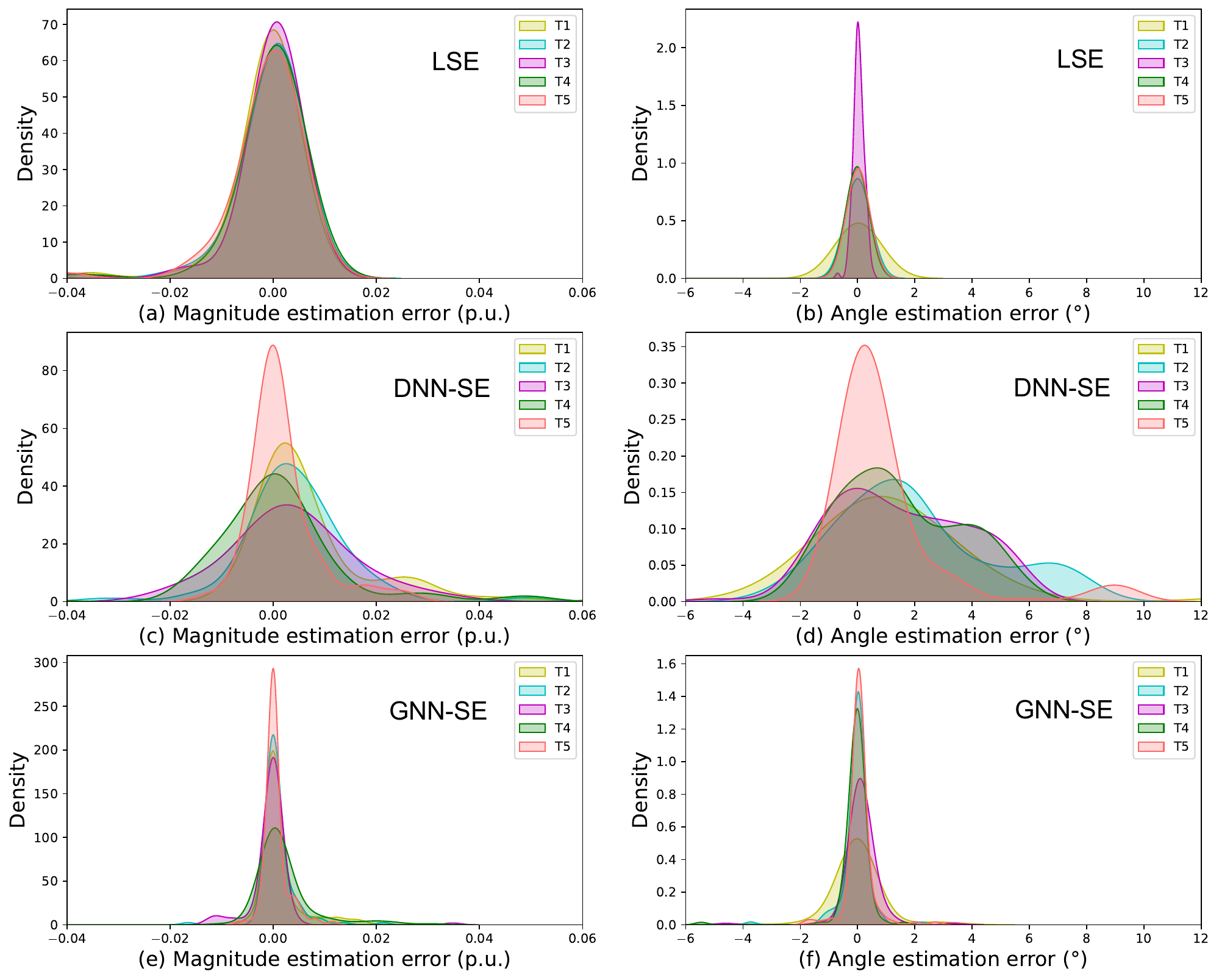}
\vspace{-2em}
\caption{Comparing density of estimation error of LSE (plots in the top row), DNN-SE (plots in the middle row), and proposed GNN-SE (plots in the bottom row) for outages of five lines of the IEEE 118-bus system that have the highest power flowing through them; (a), (c) and (e) compare the magnitudes, while (b), (d) and (f) compare the angles, considering non-Gaussian measurement noise.}
\label{figure2}
\end{figure*}

To better visualize the estimation results associated with the five highest power-flowing line outages (denoted by ${\textbf{T}}$), we depicted the distribution of sample estimation error separately for voltage magnitude and phase angle for LSE, DNN-based SE, and GNN-based SE in Figure \ref{figure2} considering non-Gaussian measurement noise. 
In Figure \ref{figure2}, (a), (c), and (e) show the voltage magnitude estimation error distributions, while (b), (d), and (f) show the estimation error distribution for phase angles for the LSE, DNN-SE, and GNN-SE, respectively. 
The figure indicates that the LSE and DNN-SE fail to provide consistent estimates since the distributions of sample estimation errors for  magnitudes (for both LSE and DNN-SE) and angles (for DNN-SE) spread out over the horizontal axis.
Conversely, for the proposed GNN-SE, the corresponding error distributions overlap one another to a greater extent (lie within a narrower range of the horizontal axis) indicating that the proposed approach is relatively immune to topology changes as well as non-Gaussian measurement noise.
Lastly, note that as subplots (e) and (f) correspond to the outages of the highest power-flowing lines, they depict the worst performance that the proposed approach will have when a single line outage takes place in the IEEE 118-bus system. 

\section{Conclusion and Future Work}\label{Conclusions}
In this paper, a time-synchronized GNN-based SE framework is presented to estimate the static states of the power system.
The framework comprises a combination of  GCN and GAT layers that consider both feature-based information and structural information (physical connections) of the system.
The proposed approach is compared with LSE and a regular (vanilla) DNN-based SE in presence of non-Gaussian measurement noise and topology changes.
The results indicate that by using convolutional and attention layers, the GNN trained for the base topology is able to perform robust estimation under different system and sensing conditions without any \textit{ex post facto} learning.

The application of GNNs to power systems is still in its infancy,
and hence, there is scope for improvement and exploration of new areas. In the future, the following research directions will be pursued:

\begin{itemize}
    \item Feature augmentation techniques will be explored to reduce the number of PMUs needed for satisfying the fully-observed node feature matrix requirement of GNNs. This will involve investigating how existing data can be leveraged to extract more features in order to reduce the need for additional PMUs.
    \item Research will be done to improve the accuracy of the GNN-based estimator to enable it to handle more challenging scenarios, such as bad/missing PMU data as well as addition of more PMUs into the system.
    \item While much of the current research in GNNs has been focused on transmission systems, there is a need to explore GNNs' applicability to distribution systems. This will require development of new GNN architectures that can handle the unique characteristics of distribution systems (e.g., unbalance, single-phase laterals).
\end{itemize}



\bibliographystyle{ieeetr}
{\footnotesize
\bibliography{References.bib}}


\end{document}